\documentclass[11pt]{article}

\newcommand{\reseteqnum}{\setcounter{equation}{0}}

\newcommand{\bbone}{{\mathchoice {\rm 1\mskip-4mu l} {\rm 1\mskip-4mu l}
{\rm 1\mskip-4.5mu l} {\rm 1\mskip-5mu l}}}

\begin{document}
\begin{titlepage}
\title{
\hfill
\Large 
\parbox{4cm}{\normalsize KUNS-1501\\HE(TH)~98/04 \\{\tt hep-lat/9803016}}\\
\vspace{2.5cm}
\bf 
Overlap formula for the chiral multiplet
\vspace{0.6cm}
\author{
Tatsumi Aoyama\thanks{e-mail address:
aoyama@gauge.scphys.kyoto-u.ac.jp} \ \ 
and  \ 
Yoshio Kikukawa\thanks{e-mail address:
kikukawa@gauge.scphys.kyoto-u.ac.jp} 
\\
{\normalsize\em Department of Physics, Kyoto University 
}\\
{\normalsize\em Kyoto 606-8502, Japan}
}
}
\date{\normalsize March, 1998}
\maketitle
\thispagestyle{empty}

\vspace{2cm}
\begin{abstract}
\normalsize
The vacuum overlap formalism is extended to describe the supersymmetric 
multiplet of a Weyl fermion, a complex scalar boson and 
an auxiliary field in the case without interaction, based on the fact 
that supersymmetry can be maintained upto quadratic terms by 
introducing bosonic species doublers. We also obtain a local action 
which describes the chiral multiplet and discuss its symmetry structure. 
It is shown that, besides the manifest supersymmetry, 
the action possesses a chiral symmetry of the type given 
by L\"uscher and analogous bosonic symmetries which may be regarded 
as independent infinitesimal rotations of complex phases of the scalar 
and the auxiliary fields. This implies that 
the $U(1) \times U(1)_R$ symmetry of the chiral multiplet 
can be exact on the lattice.

\end{abstract}

\end{titlepage}

\section{Introduction}
\reseteqnum

The vacuum overlap 
formula\cite{original-overlap, overlap-extensive} provides 
a well-defined lattice regularization of chiral determinant. 
It can reproduce the known features of the chiral determinant 
in the continuum theory: the one-loop effective action of the
background gauge field\cite{one-loop-effective-action} 
including the consistent anomaly\cite{consistent-anomaly}, 
topological charges and fermionic zero modes 
associated with the topologically non-trivial gauge 
fields\cite{original-overlap, overlap-extensive, index-theorem}, 
the $SU(2)$ global 
anomaly\cite{odd-dim-overlap-global-anomaly,4d-su2}, 
and so on. These properties of this formalism
allows the description of the fermion number violation on the lattice. 
Several numerical applications\cite{overlap-fermion-number-violation} 
have been performed. Their results strongly suggest that
the overlap formalism can actually be a promising building 
block for the construction of lattice chiral gauge theories. 

When applied to vector-like theories like QCD, the overlap formalism
also provides a quite satisfactory description of massless 
Dirac fermion. It has been shown by Neuberger
\cite{exact-massless-quark, more-about-exact-massless-quark}
that the vacuum overlaps for the massless Dirac fermion 
can be written as a single determinant of 
a Dirac operator which satisfies the Ginsparg-Wilson 
relation\cite{Ginsparg-Wilson-relation}. Moreover, 
L\"uscher has shown that 
the action given by the Dirac operator possesses an exact
chiral symmetry\cite{exact-chiral-symmetry}. 
For free fermion, the Dirac operator defines a local action.
The Ginsparg-Wilson relation is the clue to 
escape the Nielsen-Ninomiya theorem\cite{NN-theorem}.

The use of the overlap formalism to describe the supersymmetric 
Yang-Mills theory was also suggested in \cite{overlap-extensive}. 
It has been further considered in 
\cite{nishimura-maru, odd-dim-overlap-global-anomaly, 
more-about-exact-massless-quark}.

In this letter, we discuss an extension of the vacuum overlap
formalism to describe the chiral multiplet\cite{chiral-multiplet} 
of a single Weyl fermion and a complex scalar boson in the case 
without interaction. We first formulate a supersymmetric version 
of the domain-wall system, based on the fact that supersymmetry 
can be maintained upto quadratic terms by introducing bosonic 
species doublers. Then we derive the overlap formula 
of the partition function of the chiral multiplet.
In order to examine the symmetry structure of the theory
of the chiral multiplet so obtained, we next 
derive a local action with which a functional integral 
reproduces the partition function of the chiral multiplet.
We find that, besides the manifest supersymmetry, 
the action possesses a chiral symmetry of the type given 
by L\"uscher and analogous bosonic symmetries which may be regarded 
as independent infinitesimal rotations of complex phases of the scalar 
and the auxiliary fields. This means that all the symmetries 
of the continuum theory can be manifest on the lattice in this formalism.

\section{
Bosonic species doublers, lattice supersymmetry and the breakdown
of $U(1)\times U(1)_R$ symmetry
}
\reseteqnum

It is known that there are several difficulties in formulating
supersymmetric theories on the 
lattice\cite{dondi-nicolai,nicolai-map,
wilson-susy,staggered-susy,slac-susy,nojiri,curci-veneziano}. 
One of the difficulties is that 
the Leibniz rule breaks down on the lattice:
\begin{equation}
  \sum_n \left\{ 
\partial_\mu A_n \cdot B_n \cdot C_n 
+ A_n \cdot \partial_\mu B_n \cdot C_n 
+ A_n \cdot B_n \cdot \partial_\mu C_n 
         \right\} \not = 0 ,
\end{equation}
where $\partial_\mu$ is taken as the symmetric lattice derivative
\begin{eqnarray}
  \partial_\mu &=& \frac{1}{2} \left( \nabla_\mu + \bar\nabla_\mu
  \right) . 
\end{eqnarray}
$\nabla_\mu$ and $\bar \nabla_\mu$ are the nearest-neighbour 
forward and backward difference operators, respectively.
This breakdown of the Leibniz rule causes the difficulty in
introducing a local cubic interaction in supersymmetric manner.

On the other hand, since the lattice derivative is anti-Hermitian, the 
quadratic rule holds true, 
\begin{equation}
\sum_n \left\{ 
\partial_\mu A_n \cdot B_n + A_n \cdot \partial_\mu B_n 
\right\} = 0 .
\end{equation}
Then free lattice theories may possess manifest supersymmetry. 
In fact, it is known that, if species doublers are also introduced 
for boson, the free lattice theory of a single Weyl fermion and 
a complex scalar boson has a manifest supersymmetry. 

This fact can easily be understood by considering the cancellation
of the free energy. The action of a free Weyl fermion is given 
on a lattice by\footnote{ In our convension, $\sigma_\mu$ matrixes 
are defined by $\sigma_\mu= \left( \bbone, i\sigma_k \right)$.}
\begin{equation}
  S_F = 
\sum_n \psi_n^\dagger 
\sigma_\mu \frac{1}{2} \left( \nabla_\mu + \bar \nabla_\mu
        \right) \psi_n  .
\end{equation}
The free energy of this fermion is given by
\begin{equation}
   F_F = - \ln \left[ \prod_{p} \det i\sigma_\mu \sin p_\mu \right]
       = - \ln \left[ \prod_{p} \sin^2 p_\mu \right] .
\end{equation}
We can see that there are fifteen species doublers.
On the other hand, the action of the free complex boson is usually 
taken as the following form,
\begin{equation}
  S_B = - \sum_n \phi_n^\dagger \bar \nabla_\mu \nabla_\mu \phi_n ,
\end{equation}
and its free energy is given by
\begin{equation}
   F_B = \ln \left[ \prod_{p} 4 \sin^2 \frac{p_\mu}{2} \right] .
\end{equation}
Since there is no species doubler for boson, the bosonic free
energy cannot cancel the fermionic one. However, if 
we adopt the following action for the boson, 
\begin{equation}
  S_B = - \sum_n 
\phi_n^\dagger
\left\{ \frac{1}{2} \left( \nabla_\mu + \bar \nabla_\mu\right)\right\}^2
\phi_n ,
\end{equation}
then its free energy is given by
\begin{equation}
   F_B = \ln \left[ \prod_{p} \sin^2 p_\mu \right] .
\end{equation}
There appear fifteen species doublers for the boson and their free
energies exactly cancel the free energies of species doublers 
of fermion, as long as both the fermion and the boson are subject to
the same boundary condition.\footnote{In the following discussion, we 
will assume that this is the case.} 

Thus we are led to consider 
the lattice theory whose action is given by 
\begin{equation}
  S_0 = 
\sum_n \left\{
\bar \psi_n
\sigma_\mu \frac{1}{2} \left( \nabla_\mu + \bar \nabla_\mu\right) 
\psi_n 
-\phi_n^\ast
\left\{ \frac{1}{2} \left( \nabla_\mu + \bar \nabla_\mu \right)
\right\}^2 \phi_n 
- F_n^\ast F_n \right\}.
\end{equation}
The supersymmetry transformation is then defined by 
\begin{eqnarray}
\label{super-transformation}
\delta \phi_n &=& - \sqrt{2} \, \xi^T \epsilon \psi_n , \\
\delta \phi_n^\ast &=& + \sqrt{2} \, \bar \psi_n \epsilon \, \bar \xi^T , \\
\delta \psi_n &=& \sqrt{2} \, \sigma_\mu^\dagger \, \epsilon 
\, \bar \xi^T 
\frac{1}{2} \left( \nabla_\mu + \bar \nabla_\mu\right)\phi_n 
+ \sqrt{2} \, \xi  F_n , \\
\delta \bar \psi_n &=& \sqrt{2} \, 
 \frac{1}{2} \left( \nabla_\mu + \bar \nabla_\mu\right)\phi_n^\ast \, 
\xi^T \epsilon \, \sigma_\mu^\dagger  - \sqrt{2} \, \bar \xi F_n^\ast ,\\
\delta F_n &=& - \sqrt{2} \, \bar \xi \sigma_\mu 
\frac{1}{2} \left( \nabla_\mu + \bar \nabla_\mu\right) \psi_n ,\\
\delta F_n^\ast &=& - \sqrt{2} \, 
\frac{1}{2} \left( \nabla_\mu + \bar \nabla_\mu\right) \bar \psi_n
\sigma_\mu  \xi .
\end{eqnarray}

The degeneracy due to the species doublers of both the fermion and the 
boson can be resolved by adding the Majorana-type Wilson mass term 
\begin{equation}
  M_w = \left(m_0 - \frac{1}{2} \nabla_\mu \bar \nabla_\mu \right) ,
\end{equation}
in a supersymmetric manner:
\begin{eqnarray}
\label{Majorana-Wilson-chiral-multiplet}
S_{MW} &=&
\sum_n \left\{
\bar \psi_{n}
\sigma_\mu \frac{1}{2} \left( \nabla_\mu + \bar \nabla_\mu\right) 
           \psi_{n}
-\phi_{n}^\ast
\left\{ \frac{1}{2} \left( \nabla_\mu + \bar \nabla_\mu \right)
\right\}^2 \phi_{n}
- F_{n}^\ast F_{n} \right\}
\nonumber\\
&+& \sum_n \frac{1}{2} \left\{
\psi_{n}^T \epsilon  \, M_w \psi_{n}
+ \phi_{n} M_w F_{n}
+ F_{n} M_w \phi_{n} 
\right.\nonumber\\
&&\left. \qquad \qquad \qquad
-\psi_{n}^\dagger \epsilon  \, M_w \psi_{n}^\ast 
+ \phi_{n}^\ast  M_w F_{n}^\ast 
+ F_{n}^\ast  M_w \phi_{n}^\ast 
\right\} .
\end{eqnarray}

Note, however, that because of the introduction of the Majorana-Wilson term,
the chiral symmetry of the fermion 
\begin{equation}
  \psi_n \longrightarrow \exp( i \alpha ) \psi_n , 
\end{equation}
and the symmetry associated with the coherent rotation of the complex phases 
of the scalar and the auxiliary fields 
\begin{eqnarray}
  \phi_n &\longrightarrow& \exp( i \beta ) \phi_n , \\
   F_n   &\longrightarrow& \exp( i \beta ) F_n , 
\end{eqnarray}
are lost. This means that $U(1) \times U(1)_R$ symmetry of the chiral
multiplet cannot be maintained by this procedure.
In the following, we will formulate the same theory 
through the vacuum overlap formalism and will find that 
it can maintain all the required symmetries.

\section{Supersymmetric domain-wall system}
\reseteqnum

Since the domain-wall fermion\cite{original-kaplan} can be regarded 
as a collection of an infinite flavors of the Wilson fermion with 
a specific mass matrix\cite{pre-overlap,almost-massless-fermion},
it is then rather straightforward to formulate 
the supersymmetric version of the domain-wall system. The action may 
be written in the form
\begin{eqnarray}
S &=&
a \sum_{ns} \left\{
\bar \psi_{1ns}
\sigma_\mu \frac{1}{2} \left( \nabla_\mu + \bar \nabla_\mu\right) 
           \psi_{1ns}
-\phi_{1ns}^\dagger
\left\{ \frac{1}{2} \left( \nabla_\mu + \bar \nabla_\mu \right)
\right\}^2 \phi_{1ns}
- F_{1ns}^\ast F_{1ns} \right\}
\nonumber\\
&+&
a \sum_{ns} \left\{
\bar \psi_{2ns}
\sigma_\mu \frac{1}{2} \left( \nabla_\mu + \bar \nabla_\mu\right) 
           \psi_{2ns}
-\phi_{2ns}^\dagger
\left\{ \frac{1}{2} \left( \nabla_\mu + \bar \nabla_\mu \right)
\right\}^2 \phi_{2ns}
- F_{2ns}^\ast F_{2ns} \right\}
\nonumber\\
&+&
a\sum_{nst} \left\{
\psi_{1ns }^T \epsilon  \, M(s,t) \psi_{2nt}
+ \phi_{1ns} M(s,t) F_{2nt}
+ F_{1ns} M(s,t) \phi_{2nt} 
\right.\nonumber\\
&&\left. \qquad \qquad
- \bar \psi_{2ns} \epsilon  \, M^\dagger(s,t) \bar \psi_{1nt}^T
+ \phi_{2ns}^\ast  M^\dagger(s,t) F_{1nt}^\ast 
+ F_{2ns}^\ast  M^\dagger(s,t) \phi_{1nt}^\ast 
\right\} , \nonumber\\
\end{eqnarray}
where 
\begin{equation}
  M(s,t) = 
\left( - m_0 \, {\rm sign}(s+\frac{1}{2}) 
 - \frac{1}{2} \nabla_\mu \bar \nabla_\mu \right) \delta_{st}
  + \frac{1}{a} \left( \delta_{st} -\delta_{s+1,t}\right) .
\end{equation}
We only keep the lattice spacing of the fifth 
dimension, which we denote by $a$, for the convenience 
to deduce the Hamiltonian.
Hereafter we will suppress the indexes for four dimensional 
lattice, $n,m$.  

\section{Transfer matrix of the bosonic sector}
\reseteqnum

The transfer matrix of the bosonic part can be obtained by the 
standard method, which is quite analogous to the fermionic case given 
by Neuberger and Narayanan in \cite{original-overlap}.
It is also possible to obtain it starting from the vector-like 
formulation\cite{vector-like-domain-wall-fermion},  
which may be truncated at finite flavors 
just as discussed by Neuberger in \cite{almost-massless-fermion}. 
Here we follow the latter method for simplicity. We will summarize
the derivation of the transfer matrix by the former method 
in the appendix~A.

Following \cite{almost-massless-fermion}, instead of considering 
the domain-wall system, we simply take the $k$ flavor pairs of free 
chiral multiplets from the negative mass region of $-m_0$. 
We introduce the bosonic field variable defined as 
\begin{equation}
  \Phi_{s} = \left( 
\begin{array}{c} 
F_{2s} \\ \phi_{2s} \\ \phi_{1s}^\ast \\ F_{1s}^\ast 
\end{array}
\right) .
\end{equation}
Then the bosonic part of the action may be written as 
\begin{equation}
  S_B = \Phi_{s}^\dagger {\cal D}_B(s,t) \Phi_{t} ,
\end{equation}
where 
\begin{eqnarray}
{\cal D}_B=\left( 
\begin{array}{ccccccccc}
\sigma_1 \tilde C \sigma_1 & B \bbone & 0 & 0 & 0 & 0 &\cdots &  0 & 0 \\
B \bbone &  \tilde C & -1 & 0 & 0 & 0 &\cdots &  0 & 0 \\
0& -1 & \sigma_1 \tilde C \sigma_1 & B \bbone & 0 & 0 &\cdots &  0 & 0 \\
0& 0 & B \bbone &  \tilde C & -1 & 0 &\cdots &  0 & 0 \\
0&0& 0& -1 & \sigma_1 \tilde C \sigma_1 & B \bbone &\cdots & 0 & 0 \\
0&0& 0& 0 & B \bbone &  \tilde C & \cdots &0 & 0 \\
\vdots & \vdots & \vdots & \vdots & \vdots & \vdots & \ddots & \vdots 
& \vdots \\
0 & 0 & 0 & 0 & 0 & 0 & \cdots & \sigma_1 \tilde C \sigma_1 & B \bbone \\
0 & 0 & 0 & 0 & 0 & 0 & \cdots & B \bbone &  \tilde C 
\end{array} \right) . \nonumber\\
\end{eqnarray}
This is the $4k\times 4k$ matrix-valued operator acting on the 
lattice index, which we have suppressed for simplicity.
$\tilde C$ is a two-by-two matrix defined by 
\begin{eqnarray}
  \tilde C = \left( \begin{array}{cc} 
- \left\{ \frac{1}{2} \left( \nabla_\mu + \bar \nabla_\mu \right)
\right\}^2
& 0 \\
0 & -1 \end{array} \right) . 
\end{eqnarray}
$\bbone$ is a two-by-two unit matrix and the factor $B$ 
is defined by
\begin{equation}
  B = 1 
+ a \left( - m_0 
           - \frac{1}{2} \nabla_\mu \bar \nabla_\mu \right)  .           
\end{equation}

The partition function of the bosonic part reads,
\begin{equation}
  Z_B = \int \prod_s {\cal D}\Phi_s {\cal D}\Phi_s^\dagger 
        \, \exp\left( - S_B \right) 
      = \left( \det {\cal D}_B \right)^{-1} .
\end{equation}
In evaluating the determinant, we can apply the formula given 
in the appendix of \cite{almost-massless-fermion}. The result is 
\begin{equation}
  \det {\cal D}_B = (-)^{qk} \left( \det B \bbone \right)^k 
  \det\left[ 1+ e^{ k a H_B} \right]
  \det \left[ \frac{1+\Gamma_5 \tanh \left(\frac{k}{2} a H_B\right) }{2}
       \right] ,
\end{equation}
where the transfer matrix is identified as 
\begin{eqnarray}
e^{- a H_B }&=& \left( 
  \begin{array}{cc} \frac{1}{B} \bbone & - a \frac{1}{B} \tilde C \\
                    a \sigma_1 \tilde C \sigma_1 \frac{1}{B} &
   - a^2 \sigma_1 \tilde C \sigma_1 \frac{1}{B} \tilde C + B \bbone
  \end{array} \right) ,
\end{eqnarray}
\begin{equation}
  \Gamma_5 
= \left( \begin{array}{cc} \bbone & 0 \\ 0 & - \bbone \end{array}
\right) .
\end{equation}

The result of the fermionic sector has been obtained by Neuberger in 
\cite{almost-massless-fermion}:
\begin{equation}
  \det {\cal D}_F = (-)^{qk} \left( \det B \bbone \right)^k 
   \det\left[ 1+ e^{ k a H_F} \right]
  \det \left[ \frac{1+\gamma_5 \tanh \left(\frac{k}{2} a H_F\right) }{2}
       \right] ,
\end{equation}
where the transfer matrix is identified as 
\begin{eqnarray}
e^{- a H_F }&=& \left( 
  \begin{array}{cc} \frac{1}{B} \bbone & a \frac{1}{B} C \\
                    a C^\dagger \frac{1}{B} &
    a^2 C^\dagger \frac{1}{B} C + B \bbone
  \end{array} \right) ,
\end{eqnarray}
and
\begin{equation}
  C = \sigma_\mu 
      \frac{1}{2}\left( \nabla_\mu + \bar \nabla_\mu \right) .
\end{equation}

Because of the supersymmetry, this determinant of the fermionic part
is identical to the determinant of the bosonic part:
\begin{equation}
  \det {\cal D}_F = \det {\cal D}_B .
\end{equation}
This is also true if we take the anti-periodic boundary condition 
in the fifth-dimension(the flavor-space), which implies that 
\begin{equation}
(-)^{qk} \left( \det B \bbone \right)^k 
  \det\left[ 1+ e^{ k a H_F} \right]  
=(-)^{qk} \left( \det B \bbone \right)^k 
  \det\left[ 1+ e^{ k a H_B} \right] .
\end{equation}
Therefore, the supersymmetric relation implies that 
the total partition function of the chiral multiplets is unity, 
but it may be written as
\begin{equation}
  Z = \det \left[ \frac{1+\gamma_5 \tanh \left(\frac{k}{2} a H_F\right) }{2}
       \right]
\frac{1}
{  \det \left[ \displaystyle 
              \frac{1+\Gamma_5 \tanh \left(\frac{k}{2} a H_B\right)}
              {\displaystyle 2}
       \right] } .
\end{equation}

Taking the limit of the infinite extent of the fifth 
dimension(the infinite number of flavors) and the continuum 
limit in the fifth-dimension(the flavor-space), we obtain a 
formula for the partition function of the ``vector-like'' pair 
of the chiral multiplets: 
\begin{equation}
\label{overlap-vector-like-chiral-multiplet}
Z=
\det \left[ \frac{ 1+ \gamma_5 \epsilon\left(H_F \right) } {2}
       \right]
\frac{1}
{  \det \left[ \frac{\displaystyle 1+\Gamma_5 \epsilon\left( H_B
      \right) } {\displaystyle 2}
       \right] } ,
\end{equation}
where
\begin{equation}
  \epsilon \left( H \right) = \frac{H}{\sqrt{H^2}} ,
\end{equation}
\begin{eqnarray}
  H_F&=& \left( 
         \begin{array}{cc} 
  \left(-m_0 -\frac{1}{2}\nabla_\mu \bar \nabla_\mu \right)\bbone & 
  C \\
  C^\dagger & 
 -\left(-m_0 -\frac{1}{2}\nabla_\mu \bar \nabla_\mu \right) \bbone
         \end{array}
         \right) ,  \\
  H_B&=& \left( 
         \begin{array}{cc} 
  \left(-m_0 -\frac{1}{2}\nabla_\mu \bar \nabla_\mu \right)\bbone & 
  \tilde C \\
 -\sigma_1 \tilde C \sigma_1 & 
 -\left(-m_0 -\frac{1}{2}\nabla_\mu \bar \nabla_\mu \right) \bbone
         \end{array}
         \right) .
\end{eqnarray}
Note also that 
\begin{equation}
H_F^2 = H_B^2 = X^\dagger X 
                \left( \begin{array}{cc} \bbone & 0 \\ 
                                         0 & \bbone \end{array}
                \right),
\end{equation}
\begin{equation}
X^\dagger X =
\left[
-\left\{\frac{1}{2}\left( \nabla_\mu + \bar \nabla_\mu \right)\right\}^2 
+ \left( -m_0 -\frac{1}{2} \nabla_\mu \bar\nabla_\mu \right)^2 
\right] .
\end{equation}

\section{Overlap formula for the chiral multiplet}
\reseteqnum

Next we discuss the factorization property of 
the partition function of the ``vector-like'' pair of the chiral
multiplets given by Eq.~(\ref{overlap-vector-like-chiral-multiplet})
into overlaps. We follow the argument given 
by Narayanan\cite{from-Ginsparg-Wilson-relation-to-overlap}.
$H_F$ is an hermitian matrix and can be diagonalized by a certain
unitary matrix:
\begin{equation}
H_F U 
= U \, 
\left( \begin{array}{cc} \lambda_+ & 0 \\ 0 & \lambda_- \end{array}
 \right) , 
\end{equation}
where $\lambda_\pm$ are the diagonal matrixes which consist of the
positive eigenvalues and negative eigenvalues of $H_F$, 
respectively. We denote $U$ in the chiral basis as 
\begin{equation}
U = 
\left( \begin{array}{cc} U_{R+} & U_{R-} \\ 
                         U_{L+} & U_{L-} \end{array} \right) .
\end{equation}
For the free theory, it is possible to fix the phases of the
eigenvectors so that the determinant of $U$ would be unity.
\begin{equation}
  \det U = 1 .
\end{equation}
Using $U$, the Dirac operator can be written as 
\begin{equation}
\frac{\displaystyle 1+\gamma_5 \epsilon\left( H_F
      \right) } {\displaystyle 2}
= 
\left( \begin{array}{cc}  U_{R+} & 0 \\ 0 &  U_{L-} \end{array} \right) \,
U^\dagger  ,
\end{equation}
and we obtain
\begin{equation}
\det \left[ 
\frac{\displaystyle 1+\gamma_5 \epsilon\left( H_F
      \right) } {\displaystyle 2}
     \right] 
= \det U_{R+} \det U_{L-} .
\end{equation}

In the bosonic case, $H_B$ is real non-symmetric matrix. 
But, it can be diagonalized by a certain matrix, 
\begin{equation}
H_B \, O
= O \, 
\left( \begin{array}{cc} \lambda_+^\prime & 0 \\ 
                         0 & \lambda_-^\prime \end{array}
 \right) , 
\end{equation}
where $\lambda_\pm^\prime$ are the diagonal matrixes which consist of the
positive eigenvalues and negative eigenvalues of $H_B$, respectively.
It is also possible to normalize the eigenvectors so that
the determinant of $O$ would be unity.
\begin{equation}
  \det O = 1 .
\end{equation}
If we denote $O$ in the ``chiral basis'' with respect to $\Gamma_5$ as 
\begin{equation}
O = 
\left( \begin{array}{cc} O_{R+} & O_{R-} \\ 
                         O_{L+} & O_{L-} \end{array} \right) ,
\end{equation}
we have
\begin{equation}
\frac{\displaystyle 1+\Gamma_5 \epsilon\left( H_B
      \right) } {\displaystyle 2}
= 
\left( \begin{array}{cc}  O_{R+} & 0 \\ 0 &  O_{L-} \end{array} \right) \,
O^{-1} .
\end{equation}
Therefore we obtain
\begin{equation}
\det \left[ \frac{\displaystyle 1+\Gamma_5 \epsilon\left( H_B
      \right) } {\displaystyle 2}
       \right] 
= \det O_{R+} \det O_{L-} .
\end{equation}

From these factorization properties, we may define the partition
function of the single chiral multiplet by the following overlaps:
\begin{equation}
  Z_{chiral} = \det U_{L+} \, \frac{1}{\displaystyle \det O_{L+} } .
\end{equation}
In fact, if $U$ and $O$ are chosen as given in the appendix~B, 
the overlaps are evaluated as 
\begin{eqnarray}
\det U_{L+} = \det O_{L+} = \det \left[ 
\frac{( \sqrt{X^\dagger X} -m_0-\frac{1}{2}\nabla \bar \nabla)}
{2 \sqrt{X^\dagger X}} 
\right]
\end{eqnarray}
and the supersymmetric relation holds true.

\section{Action of the chiral multiplet}
\reseteqnum

We will next discuss a local action which describes the chiral
multiplet. It can be obtained by reducing the half degrees 
of freedom of the 
field variables in the vector-like theory with Majorana condition.
Once the local action is given, we can discuss the symmetry structure 
of the theory. 

The partition function of the vector-like pair of the chiral
multiplets can be expressed as a functional integral based 
with a local action, 
\begin{equation}
  Z = \int {\cal D}\Psi {\cal D}\bar \Psi 
           {\cal D}\Phi {\cal D} \Phi^\dagger 
           \, \exp\left( - \bar \Psi D_F \Psi 
                         - \Phi^\dagger D_B \Phi \right)  ,
\end{equation}
where
\begin{eqnarray}
  D_F&=&  1+ \gamma_5 H_F \frac{1}{\sqrt{X^\dagger X} } , \\
  D_B &=& 1+ \Gamma_5 H_B \frac{1}{\sqrt{X^\dagger X} } .
\end{eqnarray}
As shown by Neuberger, $D_F$ satisfies the Ginsparg-Wilson 
relation,
\begin{equation}
  D_F \gamma_5 + \gamma_5 D_F = D_F \gamma_5 D_F .
\end{equation}
Furthermore, as shown by L\"uscher, the fermionic action possesses
the symmetry under the chiral transformation
\begin{equation}
\label{chiral-transformation}
   \delta \Psi = \gamma_5 \left( 1- \frac{D_F}{2} \right) \Psi, \quad
   \delta \bar \Psi = \bar \Psi \left( 1- \frac{D_F}{2} \right) \gamma_5 .
\end{equation}

It is interesting to note that the bosonic operator $D_B$
satisfies an analogous relation, 
\begin{equation}
  D_B \Gamma_5 + \Gamma_5 D_B = D_B \Gamma_5 D_B .
\end{equation}
Accordingly, the bosonic action is symmetric under the transformation
\begin{equation}
\label{bosonic-chiral-transformation}
   \delta \Phi = \Gamma_5 \left( 1- \frac{D_B}{2} \right) \Phi, \quad
   \delta \Phi^\dagger 
 = \Phi^\dagger \left( 1- \frac{D_B}{2} \right) \Gamma_5 .
\end{equation}

The action of the chiral multiplet can be obtained by reducing
the half degrees of freedom of the field variables in 
$\Phi$ and $\Psi$ with Majorana condition. This means that
we may express $\Phi$ and $\Psi$ in terms of a two-component Weyl 
spinor, a complex scalar and an auxiliary field as 
\begin{eqnarray}
\Psi_n &=& \frac{1}{\sqrt{2}} 
\left( \begin{array}{c} - \epsilon \bar \psi_n^T \\
                                \psi_n \end{array} \right),  \quad
\bar \Psi_n = \frac{1}{\sqrt{2}} 
\left( \begin{array}{cc} 
                    \bar \psi_n & \psi_n^T \epsilon \end{array}  \right) ,
\\
\Phi_n &=&  \frac{1}{\sqrt{2}} 
\left( \begin{array}{c} F_n \\ \phi_n \\ 
                                     \phi_n^\ast \\ F_n^\ast
                                   \end{array} \right) , \quad
\Phi_n^\dagger =  \frac{1}{\sqrt{2}} 
\left( \begin{array}{cccc} \phi_n & F_n & 
                           F_n^\ast & \phi_n^\ast
                                   \end{array} \right) .
\end{eqnarray}
Then the above action reduces to 
\begin{eqnarray}
S&=& \sum_n \left\{ 
\bar \psi_n \sigma_\mu 
      \frac{1}{2}\left( \nabla_\mu+\bar\nabla_\mu \right) 
      \frac{1}{\sqrt{ X^\dagger X }} \, \psi_n  
\right.
\nonumber\\
&& \left. \qquad
- \phi_n^\ast 
\left\{\frac{1}{2}\left( \nabla_\mu + \bar \nabla_\mu \right)\right\}^2 
\frac{1}{\sqrt{ X^\dagger X } } \, \phi_n 
- F_n^\ast \frac{1}{\sqrt{ X^\dagger X }} \, F_n
\right\}
\nonumber\\
&+& \sum_n 
\frac{1}{2} \left\{
\psi_n^T \epsilon \left( 1 + 
\frac{M_w^-}{\sqrt{ X^\dagger X }}
\right) \psi_n 
+ \phi_n 
\left( 1 + 
\frac{M_w^-}{\sqrt{ X^\dagger X }} \right) F_n 
+ F_n 
\left( 1 + 
\frac{M_w^-}{\sqrt{ X^\dagger X }} \right) \phi_n 
\right.
\nonumber\\
&& \left.
\qquad \quad
- \bar \psi_n \epsilon \left( 1 + 
\frac{M_w^- }{\sqrt{ X^\dagger X }}
\right) \bar \psi_n^T 
+ \phi_n^\ast 
\left( 1 + 
\frac{M_w^-}{\sqrt{ X^\dagger X }} \right) F_n^\ast  
+ F_n^\ast  
\left( 1 + 
\frac{M_w^-}{\sqrt{ X^\dagger X }} \right) \phi_n^\ast 
\right\} ,
\nonumber\\
\end{eqnarray}
\begin{equation}
M_w^- = \left(- m_0 - \frac{1}{2} \nabla_\mu \bar \nabla_\mu \right) .
\end{equation}
This action should be compared to 
Eq.~(\ref{Majorana-Wilson-chiral-multiplet}). 

The remarkable point about this action is that 
it possesses as many symmetries as the target continuum theory.
First of all, it possesses the manifest supersymmetry under the 
super transformation of Eq.~(\ref{super-transformation}).
Secondly, it possesses a chiral symmetry of the type given 
by L\"uscher, which corresponds to the transformation 
Eq.~(\ref{chiral-transformation}). 
In terms of the single Weyl spinor variable, the transformation reads
\begin{eqnarray}
\label{chiral-transformation-single}
  \delta \psi_n 
&=& - \frac{1}{2} \psi_n + \frac{1}{2} \frac{1}{\sqrt{X^\dagger X }} 
\left[
 \left(-m_0 -\frac{1}{2} \nabla_\mu \bar\nabla_\mu \right) \psi_n
- \sigma_\mu^\dagger 
  \frac{1}{2}\left( \nabla_\mu + \bar\nabla_\mu \right) 
   \epsilon \bar \psi^T  
\right] , 
\nonumber\\
\delta \bar \psi_n 
&=&  + \frac{1}{2} \bar \psi_n 
-\frac{1}{2} 
\left[
 \bar \psi_n
  \left(-m_0 -\frac{1}{2} \nabla_\mu \bar\nabla_\mu \right) 
+ \psi^T \epsilon 
\sigma_\mu^\dagger 
\frac{1}{2}\left( \nabla_\mu + \bar\nabla_\mu \right) 
\right] \frac{1}{\sqrt{X^\dagger X }} .  \nonumber\\
\end{eqnarray}
Thirdly, the bosonic part of the action possesses the symmetries 
associated with independent infinitesimal rotations of complex
phases of the scalar and the auxiliary fields.
It is easily seen that the action is invariant under 
the asymmetric rotation of the complex phases of the scalar 
and the auxiliary fields:
\begin{eqnarray}
  \delta \phi_n &=& \phi_n ,  \\
  \delta F_n &=& - F_n .
\end{eqnarray}
Moreover, corresponding to the transformation 
Eq.~(\ref{bosonic-chiral-transformation}), the action possesses the
symmetry under the transformation:
\begin{eqnarray}
\label{bosonic-chiral-transformation-single}
  \delta \phi_n &=& \frac{1}{2} \phi_n 
-\frac{1}{2} \frac{1}{\sqrt{X^\dagger X }} 
                 \left[ 
\left(-m_0 -\frac{1}{2} \nabla_\mu \bar\nabla_\mu \right) \phi_n 
- F_n^\ast
                 \right] , \\
\delta F_n &=& \frac{1}{2} F_n 
-\frac{1}{2} \frac{1}{\sqrt{X^\dagger X }} 
      \left[ 
 \left(-m_0 -\frac{1}{2} \nabla_\mu \bar\nabla_\mu \right) F_n 
-\left\{\frac{1}{2}\left( \nabla_\mu + \bar \nabla_\mu \right)\right\}^2 
 \phi_n^\ast
                 \right] . \nonumber\\
\end{eqnarray}
This transformation can be regarded as the coherent infinitesimal 
rotation of the complex phases of the scalar and the auxiliary fields. 
These latter two symmetries implies that 
the $U(1) \times U(1)_R$ symmetry of the chiral multiplet 
can be exact on the lattice, if we formulate it through the 
vacuum overlap formalism.

\section{Discussion}
\reseteqnum

In summary, we have seen that the vacuum overlap formalism 
can provide a natural description of the supersymmetric 
chiral multiplet on the lattice: all the symmetries of 
the target continuum theory can be manifest in this formalism.
This feature may be useful in attempts of the construction 
of supersymmetric theories on the lattice. 

It may be interesting to formulate the chiral multiplet discussed 
in this letter in terms of the superfield\cite{superfield}. 
In particular, the structure of the lattice superspace should 
be examined. This is because the lattice counterpart of the 
$U(1)\times U(1)_R$ symmetry transformation, which consists of 
the transformations of Eqs.~(\ref{chiral-transformation-single}) and 
(\ref{bosonic-chiral-transformation-single}), 
mixes the field variables with their conjugates. This suggests
that the structure of the lattice superspace could be 
different from the continuum theory counterpart.

The next issue would be to formulate a lattice Wess-Zumino model
by constructing local and supersymmetric cubic interactions 
among chiral multiplets. The work in this direction is in progress.

\section*{Acknowledgments}
The authors would like to thank H.~Neuberger and T.~Kugo for 
enlightening discussions. 

\section*{Appendix}
\appendix
\reseteqnum

\section{Transfer matrix of the bosonic part}
In this appendix, we explain the derivation of the transfer matrix
for the bosonic part of the supersymmetric domain-wall system in 
some details. It turns out that the following holomorphic variables 
may be regarded to consist the canonical variables:
\begin{eqnarray}
z_{1s} &=& \sqrt{B_s} F_{2s}  , \\
\bar z_{1s} &=& \sqrt{B_s} F_{2s}^\ast , \\
z_{2s} &=& \sqrt{B_s} \phi_{2s} , \\
\bar z_{2s} &=& \sqrt{B_s} \phi_{2s}^\ast , 
\end{eqnarray}
and 
\begin{eqnarray}
z_{1s}^\ast &=& \sqrt{B_s} \phi_{1s} , \\
\bar z_{1s}^\ast &=& \sqrt{B_s} \phi_{1s}^\ast , \\
z_{2s}^\ast &=& \sqrt{B_s} F_{1s} , \\
\bar z_{2s}^\ast &=& \sqrt{B_s} F_{1s}^\ast ,
\end{eqnarray}
where
\begin{equation}
  B_s = 1 
+ a \left( - m_0 \, {\rm sign}(s+\frac{1}{2}) 
           - \frac{1}{2} \nabla_\mu \bar \nabla_\mu \right)  .           
\end{equation}
Then the bosonic part of the partition function can be written in the form:
\begin{eqnarray}
  Z_B 
&=& \int
{\cal D}\phi_1{\cal D}\phi_1^\dagger {\cal D}F_1{\cal D}F_1^\dagger
{\cal D}\phi_2{\cal D}\phi_2^\dagger {\cal D}F_2{\cal D}F_2^\dagger \, 
e^{-S_B} \nonumber\\
&=& \int 
\prod_s \left( \det B_s \right)^4 dZ_s dZ_s^\ast e^{-Z_s^\ast Z_s}  \, 
T_s(Z_s^\ast,Z_{s+1}) , 
\end{eqnarray}
where 
\begin{equation}
T_s(Z_s^\ast,Z_{s+1}) 
= L_s(Z_s^\ast) \, K_{s,s+1}(Z_s^\ast,Z_{s+1}) \, R_{s+1}(Z_{s+1}) ,
\end{equation}
\begin{eqnarray}
K_{s,s+1}(Z_s^\ast,Z_{s+1})&=& 
\exp\left( 
z_{1s}^\ast \frac{1}{\sqrt{B_s}} \frac{1}{\sqrt{B_{s+1}}} z_{1s+1} 
+\bar z_{1s}^\ast \frac{1}{\sqrt{B_s}} \frac{1}{\sqrt{B_{s+1}}} 
\bar z_{1s+1} 
\right. \nonumber\\
&& \qquad \qquad 
\left.
+
z_{2s}^\ast \frac{1}{\sqrt{B_s}} \frac{1}{\sqrt{B_{s+1}}} z_{2s+1} 
+\bar z_{2s}^\ast \frac{1}{\sqrt{B_s}} \frac{1}{\sqrt{B_{s+1}}} 
\bar z_{2s+1}
\right) , \nonumber\\
&& \\
L_s(Z_s^\ast)&=&
\exp\left( a \, 
\bar z_{1s}^\ast \frac{1}{\sqrt{B_s}}
\left\{ \frac{1}{2} \left( \nabla_\mu + \bar \nabla_\mu \right)
\right\}^2 \frac{1}{\sqrt{B_s}}z_{1s}^\ast
- a \, \bar z_{2s}^\ast \frac{1}{B_s} z_{2s}^\ast
\right) , \nonumber\\
&& \\
R_{s+1}(Z_{s+1})&=&
\exp\left( 
a \, 
\bar z_{2s+1} \frac{1}{\sqrt{B_{s+1}}}
\left\{ \frac{1}{2} \left( \nabla_\mu + \bar \nabla_\mu \right)
\right\}^2 \frac{1}{\sqrt{B_{s+1}}} z_{2s+1}
\right. \nonumber\\
&& \qquad \qquad \qquad \qquad \qquad \qquad \qquad 
\left.
- a \, \bar z_{1s+1} \frac{1}{B_{s+1}} z_{1s+1} 
\right) . \nonumber\\
\end{eqnarray}
Note that we have made use of the following notations:
\begin{eqnarray}
Z &=& \{ z_1, \bar z_1, z_2, \bar z_2 \} ,  \\
Z^\ast
&=& \{ z_1^\ast, \bar z_1^\ast, z_2^\ast, \bar z_2^\ast \} , 
\end{eqnarray}
and 
\begin{eqnarray}
  dZdZ^\ast e^{-Z^\ast Z} =
dz_1dz_1^\ast d\bar z_1d\bar z_1^\ast
dz_2dz_2^\ast d\bar z_2d\bar z_2^\ast \,
e^{-z_1^\ast z_1 -z_2^\ast z_2 
   -\bar z_1^\ast \bar z_1 -\bar z_2^\ast \bar z_2 } .
\end{eqnarray}

Next we introduce the bosonic operators which satisfy the canonical
commutation relation 
\begin{eqnarray}
  \left[ a_{1n}, a_{1m}^\dagger \right] &=& \delta_{nm} , \\
  \left[ \bar a_{1n}, \bar a_{1m}^\dagger \right] &=& \delta_{nm} , \\
  \left[ a_{2n}, a_{2m}^\dagger \right] &=& \delta_{nm} , \\
  \left[ \bar a_{2n}, \bar a_{2m}^\dagger \right] &=& \delta_{nm} . 
\end{eqnarray}
Using these operators, 
the transition amplitudes are translated to operators as follows:
\begin{eqnarray}
\hat K_{s,s+1}&=& 
\exp\left( 
a_{1}^\dagger \ln \left[ 
   \frac{1}{\sqrt{B_s}} \frac{1}{\sqrt{B_{s+1}}} \right] a_{1} 
+ \bar a_{1}^\dagger \ln \left[ 
   \frac{1}{\sqrt{B_s}} \frac{1}{\sqrt{B_{s+1}}} \right] \bar a_{1} 
\right. \nonumber\\
&& \qquad \qquad 
\left.
+
a_{2}^\dagger \ln \left[ \frac{1}{\sqrt{B_s}} \frac{1}{\sqrt{B_{s+1}}}
\right] a_{2} 
+\bar a_{2}^\dagger \ln \left[ 
    \frac{1}{\sqrt{B_s}} \frac{1}{\sqrt{B_{s+1}}} \right] 
\bar a_{2} \right) , \nonumber\\
&& \\
\hat L_s&=&
\exp\left( a \, 
\bar a_{1}^\dagger \frac{1}{\sqrt{B_s}}
\left\{ \frac{1}{2} \left( \nabla_\mu + \bar \nabla_\mu \right)
\right\}^2 \frac{1}{\sqrt{B_s}} a_{1}^\dagger
- a \, \bar a_{2}^\dagger \frac{1}{B_s} a_{2}^\dagger
\right) , \nonumber\\
&& \\
\hat R_s&=&
\exp\left( 
a \, 
\bar a_{2} \frac{1}{\sqrt{B_{s}}}
\left\{ \frac{1}{2} \left( \nabla_\mu + \bar \nabla_\mu \right)
\right\}^2 \frac{1}{\sqrt{B_{s}}} a_{2}
- a \, \bar a_{1} \frac{1}{B_{s}} a_{1} 
\right) . \nonumber\\
\end{eqnarray}
Then the partition function can be expressed as 
\begin{equation}
  Z_B = \langle bc + | \prod_s \left( \det B_s \right)^4
    \hat L_s \hat K_{s,s+1} \hat R_{s+1} | bc - \rangle .
\end{equation}

By introducing another set of the canonical bosonic operators defined by 
\begin{equation}
\hat b = \left( \begin{array}{c} a_1 \\ a_2 \\ \bar a_1^\dagger \\ \bar 
      a_2^\ast \end{array} \right) , \quad
\hat b^\dagger
= \left( \begin{array}{cccc} a_1^\dagger & a_2^\dagger
   & - \bar a_1 & - \bar a_2 \end{array} \right) , 
\end{equation}
we obtain
\begin{eqnarray}
\hat K_{s,s+1} &=&  \left( \det B_s \right) \left( \det B_{s+1} \right)
\exp\left( \hat b^\dagger D_s \hat b \right) \, 
\exp\left( \hat b^\dagger D_{s+1} \hat b \right) , \\
\hat L_s &=& \exp\left( \hat b^\dagger Q_L \hat b \right) , \\
\hat R_s &=& \exp\left( \hat b^\dagger Q_R \hat b \right) ,
\end{eqnarray}
where 
\begin{eqnarray}
 \exp \left( D_s \right) 
&=& \left( \begin{array}{cc} \frac{1}{\sqrt{B_s}} \bbone & 0 \\
            0 &  \sqrt{B_s} \bbone \end{array} \right) , \\
\exp\left( Q_{Ls} \right) 
&=& \left( \begin{array}{cc} 1 & 
      - a \, \frac{1}{\sqrt{B_s}} \tilde C \frac{1}{\sqrt{B_s}} \\
      0 & 1 \end{array} \right) , \\
\exp\left( Q_{Rs} \right) 
&=& \left( \begin{array}{cc} 1 & 0 \\
   a \, 
   \frac{1}{\sqrt{B_s}} \sigma_1 \tilde C \sigma_1 \frac{1}{\sqrt{B_s}}
   & 1 \end{array} \right) .
\end{eqnarray}
$\tilde C$ is a two-by-two matrix defined by 
\begin{eqnarray}
  \tilde C = \left( \begin{array}{cc} 
- \left\{ \frac{1}{2} \left( \nabla_\mu + \bar \nabla_\mu \right)
\right\}^2
& 0 \\
0 & -1 \end{array} \right) . 
\end{eqnarray}
Then we define the transfer matrix by 
\begin{equation}
\hat  T_s = \exp\left( - a \, \hat b^\dagger H_s \hat b \right) , 
\end{equation}
where
\begin{eqnarray}
\exp\left( - a H_s \right) 
&=&   \exp \left( D_s \right) 
      \exp\left( Q_{Rs} \right) \exp\left( Q_{Ls} \right) 
      \exp \left( D_s \right)  \\
&=& \left( 
  \begin{array}{cc} \frac{1}{B_s} & - a \frac{1}{B_s} \tilde C \\
                    a \sigma_1 \tilde C \sigma_1 \frac{1}{B_s} &
   - a^2 \sigma_1 \tilde C \sigma_1 \frac{1}{B_s} \tilde C + B_s
  \end{array} \right) .
\end{eqnarray}
For $s \ge 0$, the mass parameter is homogeneous and negative. We
may denote the quantities in this region with the subscript ``$-$''.
For $s \le -1$, it is homogeneous and positive. We may denote the 
quantities in that region with the subscript ``+''. The partition 
function now can be written as 
\begin{equation}
  Z_B = \prod_s \left( \det B_s \right)^2 \, 
\langle bc^\prime + | 
\prod_{s \le -1} T_+  \, \prod_{s \ge 0} T_- \, | bc^\prime - \rangle .
\end{equation}
Note that we have redefined the boundary states as 
\begin{eqnarray}
\langle bc^\prime + | &=& 
\langle bc + | \hat L_+ \exp\left( \hat b^\dagger D_+ \hat b \right) , \\
| bc^\prime - \rangle &=& 
\exp\left( \hat b^\dagger D_- \hat b \right) \, 
\hat R_- | bc - \rangle .
\end{eqnarray}

In the continuum limit in the fifth direction( $a\rightarrow 0$), 
the ``Hamiltonian'' operator is obtained as follows:
\begin{eqnarray}
\hat H_\mp &=& 
\hat b^\dagger 
\left( \begin{array}{cc}
    \left( \mp m_0  - \frac{1}{2} \nabla_\mu \bar \nabla_\mu \right) \bbone &
    \tilde C \\
    - \sigma_1 \tilde C \sigma_1 &
    - \left( \mp m_0  - \frac{1}{2} \nabla_\mu \bar \nabla_\mu \right) \bbone
      \end{array}
\right)
\hat b . \nonumber\\
\end{eqnarray}
We should note that this operator is not Hermitian. 

\section{Diagonalization of Hamiltonians}
\reseteqnum

In this appendix, we give explicit forms of the matrixes of 
$U$ and $O$, which diagonalize $H_F$ and $H_B$, respectively. 
They are given by the Fourier transforms of the following matrixes:
\begin{eqnarray}
 U&=& \frac{1}{\sqrt{2\lambda(\lambda-m_0-\frac{1}{2}\nabla \bar \nabla)}} 
    \left( \begin{array}{cc} 
      (\lambda-m_0-\frac{1}{2}\nabla \bar \nabla) \bbone & 
      C \\
      C^\dagger & 
    - (\lambda-m_0-\frac{1}{2}\nabla \bar \nabla) \bbone 
\end{array} \right) , \nonumber\\
\\
 O&=& \frac{1}{\sqrt{2\lambda(\lambda-m_0-\frac{1}{2}\nabla \bar \nabla)}} 
    \left( \begin{array}{cc} 
      (\lambda-m_0-\frac{1}{2}\nabla \bar \nabla) \bbone & 
      \tilde C \\
    - \sigma_1 \tilde C \sigma_1 & 
    - (\lambda-m_0-\frac{1}{2}\nabla \bar \nabla) \bbone 
\end{array} \right) , \nonumber\\
\end{eqnarray}
where
\begin{equation}
\lambda = \sqrt{ X^\dagger X }
= \sqrt{
-\left\{\frac{1}{2}\left( \nabla_\mu + \bar \nabla_\mu \right)\right\}^2 
+ \left( -m_0 -\frac{1}{2} \nabla_\mu \bar\nabla_\mu \right)^2 } . 
\end{equation}

\end{document}